\documentclass[11pt]{article}

\usepackage{amsthm,amsmath}
\RequirePackage[numbers]{natbib}
\RequirePackage[colorlinks,citecolor=blue,urlcolor=blue]{hyperref}

\usepackage{comment}

\usepackage{algorithm}
\usepackage{enumerate}
\usepackage{ifthen}
\usepackage[dvipsnames]{xcolor}

\usepackage[utf8]{inputenc}
\usepackage{framed}

\usepackage[utf8]{inputenc}
\usepackage{makeidx}
\usepackage{comment}
\usepackage{amsfonts}
\usepackage{amssymb}
\usepackage{booktabs}

\usepackage{gensymb}
\usepackage{tabularx}
\usepackage{float}
\usepackage{chngcntr}
\usepackage{caption}
\usepackage[labelformat=simple]{subcaption}
\usepackage{bbm}
\usepackage{lipsum}
\usepackage{enumitem}
\usepackage{hyperref}
\usepackage[english]{babel} 
\usepackage{bm} 
\usepackage[T1]{fontenc}	 
\usepackage{lmodern}
\usepackage{amsthm} 
\usepackage{graphicx} 
\usepackage{booktabs} 
\usepackage{multicol} 
\usepackage[toc,page]{appendix}

\newcommand{\footremember}[2]{%
	\footnote{#2}
	\newcounter{#1}
	\setcounter{#1}{\value{footnote}}%
}
\newcommand{\footrecall}[1]{%
	\footnotemark[\value{#1}]%
}

\allowdisplaybreaks
\numberwithin{equation}{section}

\theoremstyle{plain}

\allowdisplaybreaks

\begin{document}

\hypersetup{linkcolor=blue}

\date{\today}

\author{Ray Bai\footremember{UofSC}{Department of Statistics, University of South Carolina, Columbia, SC 29208.}\thanks{Email: \href{mailto:RBAI@mailbox.sc.edu}{\tt RBAI@mailbox.sc.edu}}, Lifeng Lin\footremember{FSU}{Department of Statistics, Florida State University, Tallahassee, FL 32306.}, Mary R. Boland\footremember{DBEI}{Department of Biostatistics, Epidemiology, and Informatics, University of Pennsylvania, Philadelphia, PA 19104.}, Yong Chen\footrecall{DBEI} \thanks{Email: \href{mailto:ychen123@upenn.edu}{\tt ychen123@upenn.edu}}  }

\title{A Robust Bayesian Copas Selection Model for Quantifying and Correcting Publication Bias 
}

\maketitle

\begin{abstract}
The validity of conclusions from meta-analysis is potentially threatened by publication bias. Most existing procedures for correcting publication bias assume normality of the study-specific effects that account for between-study heterogeneity. However, this assumption may not be valid, and the performance of these bias correction procedures can be highly sensitive to departures from normality. Further, there exist few measures to quantify the magnitude of publication bias based on selection models. In this paper, we address both of these issues. First, we explore the use of heavy-tailed distributions for the study-specific effects within a Bayesian hierarchical framework. The deviance information criterion (DIC) is used to determine the appropriate distribution to use for conducting the final analysis. Second, we develop a new measure to quantify the magnitude of publication bias based on Hellinger distance. Our measure is easy to interpret and takes advantage of the estimation uncertainty afforded naturally by the posterior distribution. We illustrate our proposed approach through simulation studies and meta-analyses on lung cancer and antidepressants. To assess the prevalence of publication bias, we apply our method to 1500 meta-analyses of dichotomous outcomes in the \textit{Cochrane Database of Systematic Reviews}.  Our methods are implemented in the publicly available \textsf{R} package \texttt{RobustBayesianCopas}.
\end{abstract}

\section{Introduction} \label{Intro}

Meta-analysis is a powerful technique for combining statistical evidence from multiple related studies. By synthesizing information from multiple studies, meta-analysis often has higher statistical power and precision than a single study \cite{JacksonRileyWhite2011}. A standard model in meta-analysis is the random effects model \cite{HigginsWhiteheadTurnerOmarThompson2001, RileyHigginsDeeks2011}, which is specified as
\begin{equation} \label{metaanalysismodel}
y_i = \theta + \tau u_i + s_i \epsilon_i, \hspace{.5cm} i=1, \ldots, n,
\end{equation}
where $u_i$ and $\epsilon_i$ are independent and distributed as $\mathcal{N}(0,1)$ for all $i=1, \ldots, n$. In \eqref{metaanalysismodel}, $y_i$ is the reported treatment effect for the $i$th study, $\theta$ is the population treatment effect of interest, the $u_i$'s are random effects for the between-study heterogeneity, and the $\epsilon_i$'s are within-study errors, further scaled by each study's reported standard error $s_i$. The parameter $\tau > 0$ quantifies the amount of between-study heterogeneity. 

The validity of meta-analysis is greatly compromised by the potential presence of publication bias, or the tendency for journals to publish studies showing significant results \cite{Whittington2004, Hopewell09}. In the presence of such bias, the studies in the published literature are a biased selection of the research in that area, resulting in biased estimation and misleading inference about $\theta$ \cite{Jackson2007}. A great deal of effort has gone into developing statistical methods to detect and correct publication bias. Graphical methods, such as the funnel plot, are one popular approach to this problem. Funnel plots assess the asymmetry of a scatter plot of the treatment effects from individual studies against their corresponding precisions. The presence of asymmetry in a funnel plot indicates potential publication bias \cite{EggerSmithSchneiderMinder1997,SterneEggerSmith2001,SterneGavaghanEgger2000}. Statistical tests to formally detect scatter plot asymmetry, such as regression tests \cite{BeggMazumdar1994,EggerSmithSchneiderMinder1997,SterneEggerSmith2001,MacaskillWalterIrwig2001} and rank-based tests \cite{DuvalTweedie2000}, have also been introduced. 

As an alternative to graph-based methods, selection models have also been developed. Unlike graphical methods, selection models directly address the issue of missing studies. The idea behind this approach is that the observed sample of studies is a biased sample of research done in a particular area, which was produced by a specific selection process. For example, \citet{Hedges1992}, \citet{GivensSmithTweedie1997}, and \citet{Rufibach2011} modeled the likelihood of a study being published as a function of the $p$-value obtained under the hypothesis that there is not a significant treatment effect. Copas and colleagues further introduced a flexible framework in which the probability of publication is modeled as a function of \textit{both} the effect size and its standard error \cite{CopasLi1997,Copas1999,CopasShi2000,CopasShi2001}. In addition to correcting for potentially biased estimates of $\theta$ in \eqref{metaanalysismodel}, several statistical tests for the Copas selection model have been developed, e.g. goodness of fit tests \cite{CopasShi2000} and score tests \cite{DuanPiaoMarksChuNingChen2019}. A detailed description of the Copas selection model is provided in Section \ref{CopasSelectionModel}. 

There is strong empirical evidence in support of using the Copas selection model for correcting publication bias. Based on 157 meta-analyses with binary outcomes, \citet{GuidoCarpenterRucker2010} showed that the Copas selection model gave superior performance over the trim-and-fill method, with better results among the 22 meta-analyses with evidence of selection bias. In spite of its benefits, the Copas selection model assumes that the study-specific random effects $u_i, i=1, \ldots, n$, are distributed as standard normal. This is actually a strong assumption, and it cannot be justified using the Central Limit Theorem, even when the number of studies is large \cite{HigginsThompsonSpiegelhalter2009,JacksonWhite2018,WangLee2019}. In Section \ref{Simulations}, we illustrate that the results for selection models that assume normally distributed random effects are highly sensitive to violations of normality. Additionally, existing non-Bayesian inference procedures for the Copas selection model rely on asymptotic arguments to construct confidence intervals for the parameters or to perform tests for publication bias \cite{CopasShi2000, NingChenPiao2017, DuanPiaoMarksChuNingChen2019}. These asymptotic approaches to inference can be potentially questionable if meta-analysis is performed with a small number of studies \cite{MathesKuss2018}. Many of the tests developed for graphical approaches or frequentist selection models may not have sufficient power for small samples. 

Given these issues, Bayesian approaches for correcting publication bias have a distinct advantage. First, Bayesian methods can provide further robustness through a robust prior on the study-specific random effects. Second, Bayesian methods automatically allow for nonasymptotic inference about unknown parameters through their marginal posterior distributions.  In this work, we introduce the \textit{robust Bayesian Copas selection} (RBC) model which specifically addresses the issue of robustness in the conventional Copas selection model. We show that modeling the study-specific effects with heavy-tailed densities can significantly reduce biased estimation of treatment effects where there is departure from normality. At the same time, our model \textit{also} performs well when the study-specific effects truly are normal. 

In standard meta-analysis \eqref{metaanalysismodel} and without addressing the issue of publication bias, a number of researchers have also raised concerns about the normality assumption for the study-specific effects (see, e.g. \cite{HigginsThompsonSpiegelhalter2009,JacksonWhite2018, WangLee2019, NegeriBeyene2020} and references therein). In \citet{NegeriBeyene2020}, the normal random effects are replaced with skew-normal distributions to provide additional modeling flexibility. In this work, we similarly place heavy-tailed priors on the $u_i$'s, except our results are placed within the context of correcting publication bias, in addition to standard meta-analysis.

Finally, we are not aware of any measures to quantify publication bias using selection models. For graphical methods, such as the funnel plot, there have been several measures proposed, including Egger's regression intercept \cite{EggerSmithSchneiderMinder1997} and the skewness of the collected studies' distribution \cite{LinChu2018}. See \citet{LinShiChuMurad2020} for a more detailed review.  However, skewness and asymmetry in funnel plots can arise from causes unrelated to study selection, such as between-study heterogeneity or the intrinsic correlations between the effect size and standard error \cite{SterneBMJ2011,DuanPiaoMarksChuNingChen2019}.

In this paper, we develop a new measure for quantifying publication bias based on the Copas selection model. Our measure quantifies the \textit{dissimilarity} between estimates obtained under the Copas selection model \eqref{Copasmodel} and those obtained under a standard meta-analysis model \eqref{metaanalysismodel}. A key benefit of our approach is that it not only quantifies the difference between two point estimates, but it \textit{also} takes into account the estimation uncertainty afforded by the full posterior distribution. Unlike the quantitative measures based on funnel plots, our measure is bounded between zero and one, with smaller values indicating negligible publication bias and values close to one indicating a very strong magnitude of publication bias. Thus, our approach has a clear and intuitive interpretation.

The rest of this paper is structured as follows. Section \ref{CopasSelectionModel} describes the Copas selection model. In Section \ref{RBCModelIntro}, we introduce the RBC model. In Section \ref{QuantificationOfBias}, we introduce the $D$ measure for quantifying publication bias based on the RBC model. In Section \ref{Simulations}, we illustrate the robustness of our approach through simulation studies. In Section \ref{DataApplication}, we apply the RBC model to meta-analyses of the relationship between second-hand tobacco smoke and lung cancer and the efficacy of antidepressants. We also assess the prevalence of publication bias using a random sample of 1500 meta-analyses of dichotomous outcomes in the \textit{Cochrane Database of Systematic Reviews}. 

\section{The Copas Selection Model} \label{CopasSelectionModel}

The Copas selection model \cite{Copas1999,CopasShi2001} is specified as follows. For all $i=1, \ldots, n$,
\begin{equation} \label{Copasmodel}
\begin{array}{rl}
y_i | z_i > 0  & = \theta + \tau u_i + s_i \epsilon_i,  \\
z_i  & = \gamma_0 + \gamma_1 / s_i + \delta_i,  \\
 \textrm{corr}(\delta_i, \epsilon_i) & = \rho, 
\end{array}
\end{equation}
where $u_i, \epsilon_i$ and $\delta_i$ are marginally distributed as $\mathcal{N}(0,1)$ and $u_i$ and $\epsilon_i$ are independent. By \eqref{Copasmodel}, the $i$th study is assumed to be published only if an associated latent variable $z_i$ (also known as the propensity score) is greater than zero. The propensity score, i.e. the propensity to publish, is characterized by two parameters $(\gamma_0, \gamma_1)$. The parameter $\gamma_0$ controls the overall likelihood of a study being published, while $\gamma_1$ characterizes how the chance of publication depends on sample size. In general, $\gamma_1$ is positive so that studies with larger sample sizes are more likely to be published. The reported effects and the propensity scores are assumed to be correlated through $\rho$, which controls how the probability of publication is influenced by the effect size of the study.  When publication bias is present, i.e. $\rho \neq 0$, standard meta-analysis will lead to biased estimation of $\theta$. On the other hand, if $\rho = 0$, then there is no publication bias, and the model \eqref{Copasmodel} reduces to the standard random effects model \eqref{metaanalysismodel}.

In \citet{CopasShi2001} and \citet{NingChenPiao2017}, the unknown parameters $(\theta, \tau, \rho)$ in \eqref{Copasmodel} are estimated using maximum likelihood estimation (MLE), conditionally on a given pair $(\gamma_0, \gamma_1)$. \citet{CopasShi2001} recommend choosing $(\gamma_0, \gamma_1)$ using a grid search. To avoid a grid search, \citet{NingChenPiao2017} assume that the biased data generating mechanism contains a latent variable that accounts for the additional unpublished studies. They treat these variables as missing data and develop an EM algorithm to simultaneously obtain the MLEs for $(\theta, \tau, \rho, \gamma_0, \gamma_1)$. 

In the Bayesian framework, \citet{MavidrisSuttonCiprianiSalanti2013} estimate the parameters in \eqref{Copasmodel} by placing priors on $(\theta, \tau, \rho)$ and informative priors on the lower and upper bounds for $\Pr(z_i > 0 | s_i)$, which act as a proxy for priors on $(\gamma_0, \gamma_1)$. To derive informative priors on the bounds for $\Pr(z_i > 0 |s_i)$, \citet{MavidrisSuttonCiprianiSalanti2013} recommend using ``both external data and an elicitation process of expert opinion.'' However, this may pose potential issues for meta-analyses when such prior information or expertise is difficult or impossible to attain. This article thus differs from the work of \cite{MavidrisSuttonCiprianiSalanti2013} in several ways. First, we place weakly informative priors on $(\gamma_0, \gamma_1)$ \textit{directly}. Second, unlike \cite{MavidrisSuttonCiprianiSalanti2013}, we also dispense with the assumption that the random effects are normally distributed. Finally, we introduce a new measure for quantifying publication bias, an issue which has not been addressed by previous works on selection models under either the Bayesian \textit{or} the frequentist framework.

\section{The Robust Bayesian Copas Selection Model} \label{RBCModelIntro}

\subsection{The RBC Prior Specification} \label{PriorSpecification}

Let $\bm{y} = (y_1, \ldots, y_n)'$, $\bm{u}= (u_1, \ldots, u_n)'$, and $\bm{z} = (z_1, \ldots, z_n)'$. Our objective is to formulate a robust Bayesian model for \eqref{Copasmodel} by placing appropriate priors on the unknown parameters.
In the RBC model, we first endow the mean effect $\theta$ in \eqref{Copasmodel} with a normal prior,
\begin{equation} \label{priorontheta}
\theta \sim \mathcal{N}(0, \sigma_{\theta}^2),
\end{equation}
where $\sigma_{\theta}^2$ is set to be a large value so that the prior on $\theta$ is fairly noninformative. As a default, we set $\sigma_{\theta}^2 = 10^4$. 

To model the between-study heterogeneity in \eqref{Copasmodel}, we follow the recommendations of \citet{Gelman2006} and \citet{PolsonScott2012} and endow the heterogeneity parameter $\tau$ with the noninformative half-Cauchy prior,
\begin{equation} \label{priorontau}
\tau \sim \mathcal{C}^{+}(0, 1).
\end{equation}
In earlier Bayesian work on the Copas selection model, \citet{MavidrisSuttonCiprianiSalanti2013} endowed the heterogeneity \textit{variance} $\tau^2$ with an inverse-gamma prior $\mathcal{IG}(a, a)$, where the hyperparameter $a$ was set to be a very small value. However, for data sets where very low values of $\tau$ are possible (i.e. when there is hardly any between-study heterogeneity), inferences can be very sensitive to the choice of $a$, and the inverse-gamma prior becomes rather informative \cite{Gelman2006,PolsonScott2012}. To circumvent this issue, we instead use the half-Cauchy prior \eqref{priorontau}.

\begin{table} 
\centering
\caption{List of priors for the study-specific effects. For the Student's $t$-distribution, $\nu > 0$ is the degrees of freedom, and in the slash distribution, $\xi > 0$ is the shape parameter. Smaller values of $\nu$ and $\xi$ lead to heavier tails. }\label{Table:1}
\resizebox{\textwidth}{!}{
\begin{tabular}{llc}
\hline
Prior & Density & Robust? \\
\hline
Standard normal & $p(u_i) = (2 \pi)^{-1/2} \exp ( - u_i^2 / 2)$ & no \\ 
Laplace & $p(u_i) = (1/2) \exp ( - \lvert u_i \rvert )$ & yes \\
Student's $t$ & $p(u_i ) \propto (1 + u_i^2 / \nu )^{- (\nu+1)/2}$ & yes \\
Slash & $p(u_i) = \xi \int_{0}^{1} \lambda^{\xi-1} \mathcal{N}( u_i \mid 0, \lambda^{-1} ) d \lambda$ & yes \\  
\hline
\end{tabular}}
\end{table} 

Next, we consider the distribution of the random effects $\bm{u}$ in \eqref{Copasmodel},
\begin{equation} \label{prioronu}
u_i \sim p(u_i), \hspace{.5cm} i=1, \ldots, n.
\end{equation}
In conventional meta-analysis, we have $u_i \sim \mathcal{N}(0, 1), i=1, \ldots, n$. If one has strong \textit{a priori} knowledge that the normality assumption holds, then our model can be implemented with standard normal priors on the random effects. As discussed earlier, however, this assumption may not be appropriate. To model study-specific effects that potentially deviate from normality, we instead endow the $u_i$'s with a heavy-tailed distribution, such as the Laplace, Student's $t$, or slash distribution.  Table \ref{Table:1} gives a list of the priors that we considered for the RBC model. 

The Laplace, Student's $t$, and slash distributions are routinely used for robust Bayesian statistical modeling \cite{LangeLittleTaylor1989, SongYaoXing2014, GeraciBottai2006, AbantoValle2010}. In this article, we extend their use to selection models for correcting publication bias in meta-analysis. In Section \ref{Simulations}, we illustrate that these distributions are more robust in the sense that when normality is violated, the heavy-tailed densities produce less biased estimates of $\theta$ than methods that assume normality \textit{a priori}. 

We note that we only (potentially) model the between-study random effects with a robust prior and not the within-study random errors $\epsilon_i, i=1, \ldots, n$. This is because there is rarely enough information in the sample of collected studies for the practitioner to model the within-study errors for any individual study. On the other hand, the sample of collected studies typically does contain enough information to model the \textit{between}-study heterogeneity.

Next, we endow the correlation parameter $\rho$ in \eqref{Copasmodel} with the noninformative uniform prior,
\begin{equation} \label{prioronrho}
\rho \sim \mathcal{U}(-1, 1).
\end{equation}

Finally, we consider the priors for $\gamma_0$ and $\gamma_1$ in \eqref{Copasmodel}, which control the probability of publication. We will ultimately place weakly informative uniform priors on these two quantities. To determine appropriate values for the hyperparameters in the priors on $(\gamma_0, \gamma_1)$, we first note that for some values $(P_{\textrm{low}}, P_{\textrm{high}} )$ between zero and one, our model should satisfy
\begin{align*}
0 \approx P_{\textrm{low}} \leq \Pr (z_i > 0 | s_i ) \leq P_{\textrm{high}} \approx 1,\hspace{.3cm}  i= 1, \ldots, n.
\end{align*}
Letting $s_{\min}$ and $s_{\max}$ denote the smallest and largest reported standard errors respectively, \citet{MavidrisSuttonCiprianiSalanti2013} showed that when $\gamma_1 \geq 0$, the above inequality translates to
\begin{align*}
\Phi^{-1} (P_{\textrm{low}}) \leq \gamma_0 + \frac{\gamma_1}{s_{\max}} \leq \gamma_0 + \frac{\gamma_1}{s_{\min}} \leq \Phi^{-1} (P_{\textrm{high}}),
\end{align*}
where $\Phi^{-1}$ denotes the inverse cumulative distribution function (cdf) for the standard normal density. Since the standard normal density places most of its mass in the interval $(-2, 2)$, this suggests the following priors for $(\gamma_0, \gamma_1)$:
\begin{equation} \label{priorongamma0}
\gamma_0 \sim \mathcal{U}(-2, 2),
\end{equation}
and
\begin{equation} \label{priorongamma1}
\gamma_1 \sim \mathcal{U}(0, s_{\max}).
\end{equation}
This ensures that most of the mass for the $z_i$'s will lie between $(-2, 3)$, leading to selection probabilities from 2.5\% to 99.7\%. As noted earlier, our prior specification \eqref{priorongamma0}-\eqref{priorongamma1} differs from the Bayesian selection model introduced by \citet{MavidrisSuttonCiprianiSalanti2013}, who placed informative priors on the lower and upper bounds of the selection probabilities $\Pr(z_i > 0 | s_i), i=1, \ldots n$. Here, we place weakly informative priors on $(\gamma_0, \gamma_1)$ \textit{directly}, with the primary aim of avoiding the need to perform problem-specific tuning of these hyperparameters. 

\subsection{Implementation and Inference Under the RBC Model}

Let $\mu_i = \theta + \tau u_i, i=1, \ldots, n$, and let $\bm{\mu} = (\mu_1, \ldots, \mu_n)'$. An alternative way to write model \eqref{Copasmodel} is 
\begin{equation} \label{bivariate}
\begin{pmatrix} y_i \\ z_i \end{pmatrix} \sim \mathcal{N}_2 \left( \begin{pmatrix} \mu_i \\ \gamma_0+\gamma_1/s_i \end{pmatrix}, \begin{pmatrix} s_i^2 & \rho s_i \\ \rho s_i & 1 \end{pmatrix} \right) \bm{1}_{z_i > 0}, \hspace{.3cm} i=1, \ldots, n,
\end{equation}
that is, $(y_i, z_i)'$ is a truncated bivariate normal distribution. Under this reparametrization, the induced prior on $\bm{\mu}$ is  $p ( \mu_i; \theta, \tau) = p ( (\mu_i - \theta ) / \tau )$, where $p( \cdot )$ is the density in \eqref{prioronu}. For example, if the random effects distribution is the Student's $t$ density $t_{\nu}$, then $\mu_i \sim t_{\nu} ( \theta, \tau), i =1, \ldots, n$, i.e. the $\mu_i$'s are generalized $t$ distributions with mean $\theta$ and scale $\tau$.

To sample from $(y_i, z_i)$ in \eqref{bivariate}, we follow the approximation introduced in \cite{MavidrisSuttonCiprianiSalanti2013}, where we first sample $z_i \sim \mathcal{N}(\gamma_0+\gamma_1/s_i, 1) \bm{1}_{z_i > 0}$, and then we sample $y_i | z_i \sim \mathcal{N} (\mu_i + \rho s_i (z_i - \gamma_0 - \gamma_1 / s_i), s_i^2(1-\rho^2))$. With this approximation for \eqref{bivariate} and the reparametrization of our model in terms of $\bm{\mu}$, we can then easily implement our model using any standard Markov chain Monte Carlo (MCMC) software by specifying the appropriate priors on  $(\bm{\mu}, \theta, \tau, \rho, \gamma_0, \gamma_1)$. Our implementation, which is available in the publicly available \textsf{R} package \texttt{RobustBayesianCopas}, uses the \texttt{JAGS} software.

Using MCMC samples, we can approximate the marginal posteriors $\pi(\theta | \bm{y})$, $\pi ( \tau | \bm{y})$, $\pi( \rho | \bm{y})$, $\pi(\gamma_0 | \bm{y})$, and $\pi(\gamma_1 | \bm{y})$. All inferences about these model parameters can then be conducted through their posterior densities.

\subsection{Selecting the Appropriate Random Effects Distribution} \label{RandomEffectsDist}

Rather than imposing a specific assumption about the distribution of the study-specific effects, we recommend fitting our model with several different distributions for the $u_i$'s and then using an appropriate criterion to select the model for the final analysis. In order to select the appropriate distribution $p ( \cdot )$ for the study-specific effects in \eqref{prioronu}, we recommend using the DIC proposed by \citet{SpiegelhalterBestCarlinVanDerLinde2002}, because of its simplicity and good empirical performance in our simulation studies. For an unknown parameter $\bm{\Xi}$, the deviance is $D(\bm{\Xi}) = - 2 \log L ( \bm{\Xi}; \textrm{Data} )$, where $L( \bm{\Xi}; \textrm{Data} )$ is the likelihood for the model. The DIC is given by $\overline{D(\bm{\Xi})} + p_D$ where $\overline{D(\bm{\Xi})}$ is the posterior mean deviance, and $p_D = \overline{D (\bm{\Xi})} - D ( \bar{\bm{\Xi}})$ is the effective number of model parameters where $D(\bar{\bm{\Xi}})$ is the deviance evaluated at the posterior mean of $\bm{\Xi}$. The DIC rewards better fitting models through $\overline{D(\bm{\Xi})}$ and penalizes more complex models through $p_D$. 

In the RBC model with the reparametrization $\mu_i = \theta + \tau u_i$ in \eqref{bivariate}, we have $\bm{\Xi} = (\bm{\mu}, \rho, \gamma_0, \gamma_1)'$, and the deviance (up to an additive constant not depending on $\bm{\Xi}$) is
\begin{align} \label{deviance}
D ( \bm{\Xi} ) = \sum_{i=1}^{n} \left[ \frac{(y_i - \mu_i)^2}{ s_i^2} + 2 \log ( \Phi ( \gamma_0 + \gamma_1 / s_i)) - 2 ( \log \Phi (v_i)) \right],
\end{align}
where $\Phi$ is the cumulative distribution function for a standard normal random variable and $v_i = \{ \gamma_0 + [ \gamma_1  + \rho (y_i - \mu_i ) ] / s_i \} / \sqrt{1 - \rho^2}$. We estimate $\overline{D ( \bm{\Xi})}$ and $\bar{\bm{\Xi}}$ using MCMC samples and then compute the DIC as 
\begin{align} \label{DIC}
\textrm{DIC} = 2  \overline{D( \bm{\Xi})} - D ( \bar{\bm{\Xi}}).
\end{align}
The model with the smallest DIC is chosen as the model with which to perform the final analysis.  

\section{Quantifying Publication Bias with the RBC Model} \label{QuantificationOfBias}

\subsection{The $D$ Measure} \label{QuantificationPosterior}

Apart from inference about $\theta$, we may be interested in quantifying the magnitude of publication bias. To the best of our knowledge, this issue has not been addressed before for selection models, although there exist several procedures based on funnel plots \cite{LinShiChuMurad2020}. Since the Copas selection model \eqref{Copasmodel} explicitly models the selection mechanism, a natural measure for quantifying publication bias is the dissimilarity between the posterior of $\theta$ under the standard random effects model \eqref{metaanalysismodel} and the posterior for $\theta$ under the RBC model where $\rho$ is \textit{also} estimated from the data. 

Throughout this section, we assume that we have already selected the density $p$ with which to model the random effects in \eqref{prioronu}, and we are now interested in quantifying the magnitude of publication bias. Let $\pi_{rbc} (\theta | \bm{y})$ be the posterior for $\theta$ under the complete RBC model with priors \eqref{priorontheta}-\eqref{priorongamma1}, Let $\pi_{\rho = 0} ( \theta | \bm{y})$ be the posterior for $\theta$ under the RBC model where we \textit{a priori} fix $\rho = 0$. When $\rho = 0$, the posterior $\pi(\theta,  \tau^2, \bm{u} | \bm{y})$ does \textit{not} depend on $(\bm{z}, \gamma_0, \gamma_1)$, and thus, it reduces to the posterior distribution under the standard random effects model \eqref{metaanalysismodel} with \textit{only} the priors \eqref{priorontheta}-\eqref{prioronu} on $(\theta, \tau^2, \bm{u})$.

 If $\rho = 0$ in \eqref{Copasmodel}, then there is no publication bias, and the estimates for $\theta$ under the standard meta-analysis model \eqref{metaanalysismodel} and the Copas selection model \eqref{Copasmodel} are theoretically the same.  Thus, the density $\pi_{\rho=0} ( \theta | \bm{y})$ can be interpreted as the posterior for $\theta$ \textit{before} bias correction. On the other hand, if $\rho \neq 0$ in \eqref{Copasmodel}, then the estimate of $\theta$ under \eqref{metaanalysismodel} is a biased estimate, whereas the estimate for $\theta$ under \eqref{Copasmodel} corrects this bias. The density $\pi_{rbc} (\theta | \bm{y})$ is therefore the posterior \textit{after} bias correction. The dissimilarity between $\pi_{\rho=0} (\theta | \bm{y})$ and $\pi_{rbc} (\theta | \bm{y})$ allows us to quantify the magnitude of publication bias.  If publication bias is negligible (i.e. $\rho \approx 0$), then the posteriors for $\theta$ before and after bias correction will be very similar to each other.  

To utilize the posterior $\pi(\theta | \bm{y})$ in our quantification of publication bias, we propose using the Hellinger distance between $\pi_{rbc} ( \theta | \bm{y})$ and $\pi_{\rho=0} (\theta | \bm{y})$. The Hellinger distance between two densities $f(x)$ and $g(x)$ is defined as
\begin{equation} \label{hellinger} 
H(f,g) = \left[ 1 - \displaystyle \int \sqrt{f(x) g (x)} dx \right]^{1/2}.
\end{equation}
Unlike other distance measures like Kullback-Leibler divergence or the total variation distance, the Hellinger distance is both symmetric \textit{and} always bounded between zero and one. The magnitude of the Hellinger distance also has a clear interpretation. Values close to zero indicate that $f$ and $g$ are nearly identical distributions, while values close to one indicate that the majority of the probability mass in $f$ does \textit{not} overlap with that of $g$. 

In the present context, we may estimate the posteriors $\pi_{rbc} ( \theta | \bm{y})$ and $\pi_{\rho=0} ( \theta | \bm{y})$ by using MCMC samples of $\theta$ to obtain kernel density estimates, $\widehat{\pi}_{rbc} (\theta | \bm{y})$ and $\widehat{\pi}_{\rho=0} ( \theta | \bm{y})$. We then use numerical integration to estimate the Hellinger distance \eqref{hellinger} between these two densities, i.e.
\begin{equation}\label{Dmeasure}
D = H \left( \widehat{\pi}_{rbc} (\theta | \bm{y}), \widehat{\pi}_{\rho=0} (\theta | \bm{y}) \right). 
\end{equation}
Smaller values of $D$ ($D \approx 0$) indicate a small or negligible magnitude of publication bias, while larger values of $D$ ($D \approx 1$) indicate a strong magnitude of publication bias. We note that $D$ can also be used to quantify the publication bias in the heterogeneity parameter $\tau$ by computing the Hellinger distance between $\pi_{rbc} ( \tau | \bm{y})$ and $\pi_{\rho = 0} ( \tau | \bm{y})$. However, since meta-analysis practitioners are mainly interested in the treatment effect, we focus on $\theta$.

\begin{figure}[t!]
	\centering
	\includegraphics[width=0.72\linewidth]{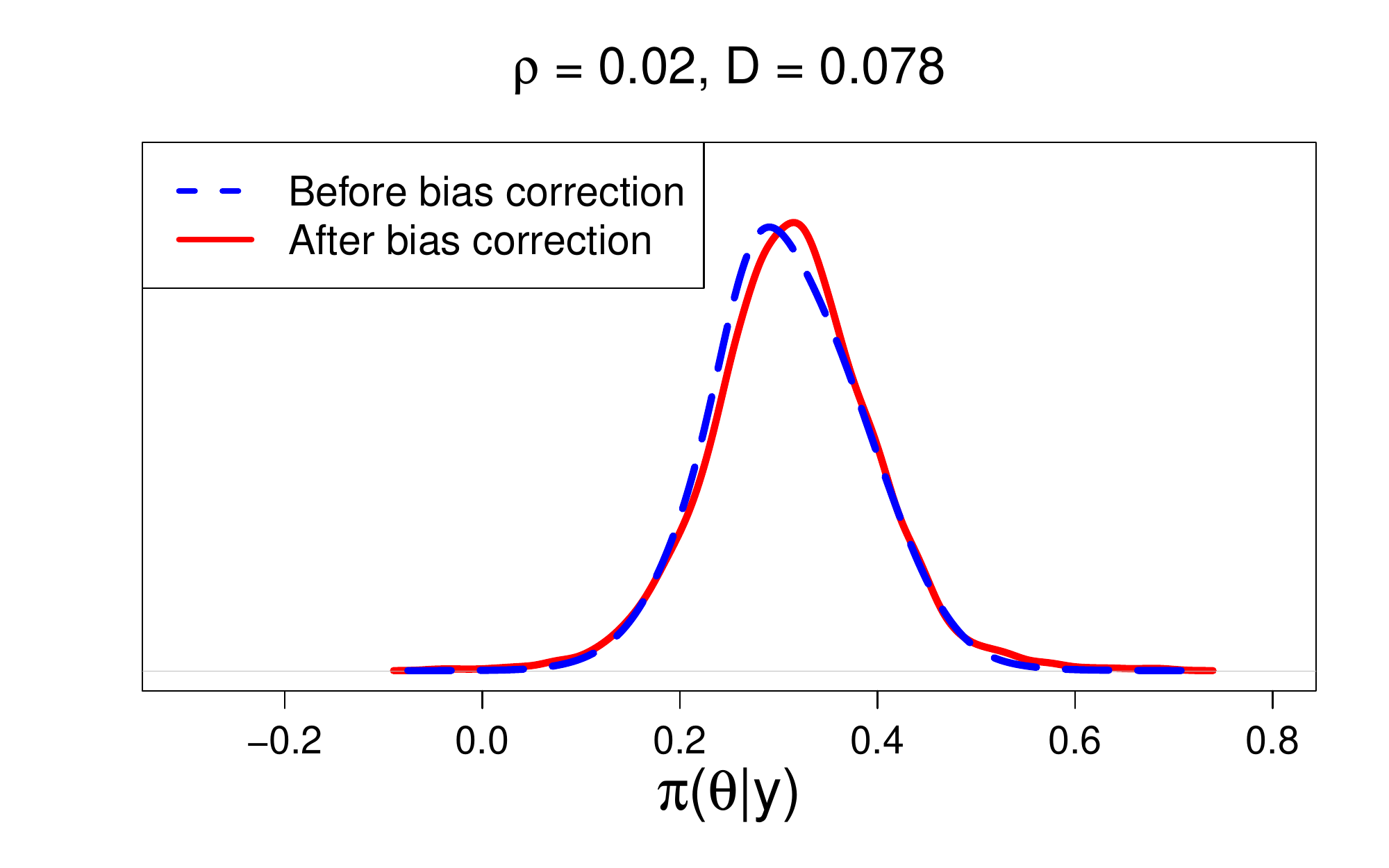}
	\includegraphics[width=0.72\linewidth]{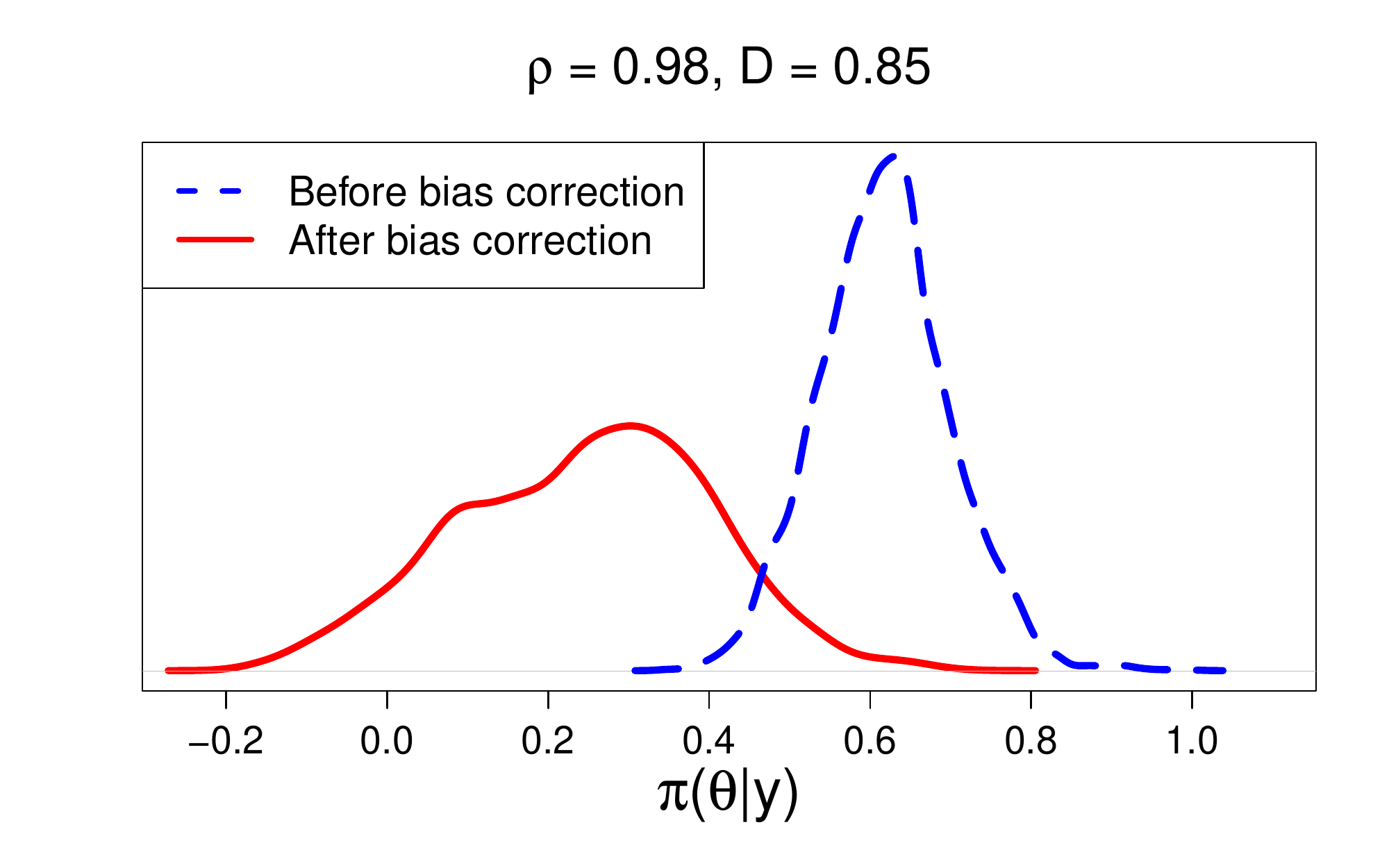}
	\caption{The posterior distributions $\pi_{rbc}( \theta | \bm{y})$ (solid line) and $\pi_{\rho=0} ( \theta | \bm{y})$ (dashed line) from two experiments where the true random effects $u_i, i=1, \ldots, n$, are distributed as $t_{3}$. In the top panel, the true $\rho = 0.02$ and $D = 0.028$. In the bottom panel, the true $\rho = 0.98$ and $D = 0.85$. }
	\label{fig:Dfs}
\end{figure}

Figure \ref{fig:Dfs} illustrates the benefits of using $D$ as a measure of publication bias magnitude. For our illustration,  we used the same settings as those from Experiment 1 in Section \ref{Simulations}, except we set $\rho=0.02$ or $\rho=0.98$. We estimated $\pi_{rbc} ( \theta | \bm{y})$ and $\pi_{\rho=0} ( \theta | \bm{y})$ using $t_4$ distributions for the random effects. In the top panel of Figure \ref{fig:Dfs}, we plot the posterior $\pi_{rbc} (\theta | \bm{y})$ (the solid line) against the posterior $\pi_{\rho=0} ( \theta| \bm{y})$ (the dashed line) when $\rho = 0.02$ (i.e. low publication bias). We see that there is significant overlap between the two distributions, and thus, $D = 0.078$. In contrast, we see in the bottom panel that when $\rho = 0.98$ (i.e. substantial publication bias), $\pi_{rbc} (\theta | \bm{y})$ and $\pi_{\rho=0} ( \theta| \bm{y})$ give more distinctive estimates of $\theta$. Moreover, $\pi_{rbc} ( \theta | \bm{y})$ has a wider spread than $\pi_{\rho=0} (\theta | \bm{y})$. This reflects the greater uncertainty about $\theta$ due to the presence of strong publication bias. When $\rho = 0.98$, we obtain $D=0.85$.

Since the magnitude of $D$ often depends on the magnitude of the correlation parameter $\rho$, one may wonder what the advantage of using $D$ is over simply using the posterior $\pi(\rho | \bm{y})$ to quantify publication bias. In our view, $D$ is a much more intuitive measure because it directly characterizes the amount of publication bias in the mean treatment effect $\theta$. As demonstrated in Figure \ref{fig:Dfs}, $D$ quantifies how much the posterior $\pi(\theta | \bm{y})$ changes \textit{after} a bias correction has been made by the RBC model.

Unlike measures based on funnel plot asymmetry, such as skewness \cite{LinChu2018}, our proposed $D$ measure does not determine the direction of the potential publication bias. However, our approach has several advantages. First, as a divergence measure between two probability distributions, $D$ automatically takes into account the estimation discrepancy \textit{and} the variability in $\theta$. Second, since $D$ is always bounded between zero and one, it has a clear interpretation. Finally, our measure quantifies the change in $\pi ( \theta | \bm{y})$ that can be solely attributed to selection bias (or how much the posterior changes because of $\rho$), whereas asymmetry may be due to factors unrelated to selection, such as methodological differences between studies \cite{SterneBMJ2011}. If the direction of the bias is of particular interest, the practitioner can examine scatter plots or the sign (positive or negative) of the sample skewness of the standardized deviates $ d_i = (y_i - \widehat{\theta}) / \sqrt{ s_i^2 + \widehat{\tau}^2 }, 1, \ldots, n$, where $(\widehat{\theta}, \widehat{\tau})$ are point estimates under the usual random effects model \eqref{metaanalysismodel} \cite{LinChu2018, MuradChuLinWang2018}.

\subsection{Interpreting the $D$ Measure} \label{Dcutoffs}

The posteriors $\pi_{rbc} ( \theta | \bm{y})$ and $\pi_{\rho=0} ( \theta | \bm{y})$ are analytically intractable and therefore have to be approximated using MCMC. Using kernel density estimation and numerical integration to evaluate the Hellinger distance \eqref{hellinger} also introduces some approximation error. Due to these approximations, we will not, in general, be able to obtain $D=0$ (which would indicate the complete absence of publication bias, or that $\pi_{rbc}(\theta | \bm{y})$ and $\pi_{\rho=0} (\theta | \bm{y})$ are identical). However, MLE approaches to estimating the parameters in \eqref{Copasmodel} and \eqref{metaanalysismodel} are also unable to produce exactly identical estimates for $\theta$, because numerical optimization must be performed on two different likelihood functions. We believe it is much more important to determine whether publication bias is \textit{negligible}, as opposed to completely absent. 

Table \ref{Dinterp} gives our suggested cutoffs for interpreting $D$. Our simulation studies and data analyses in Sections \ref{Simulations} and \ref{DataApplication} demonstrate that these guidelines seem reasonable. Nevertheless, we caution that our recommendations should be used as rough guidelines, and ``acceptable'' or ``unacceptable'' values for $D$ should be determined within the context of the problem being studied. We also recommend that practitioners examine plots of $\pi_{rbc} ( \theta | \bm{y})$ against $\pi_{\rho=0} (\theta | \bm{y})$ (such as the ones in Figures \ref{fig:Dfs} and \ref{fig:smokingantidepressants}) to visualize the extent of the bias correction made by the RBC method.

\begin{table}[t]  
	\caption{Recommended cutoffs for interpreting the $D$ measure} \label{Dinterp}
  \begin{center}
    \begin{tabular}{c|c} 
      \textbf{Range} & \textbf{Interpretation}  \\
      \hline
      $0.0 \leq D \leq 0.25$ & Negligible publication bias \\
      $0.25 < D \leq 0.5$ & Moderate publication bias\\
      $0.5 < D \leq 0.75$ & High publication bias \\
      $0.75 < D \leq 1.0$ & Very high publication bias \\
    \end{tabular}
  \end{center}
\end{table}

\section{Simulation Study for the RBC Model} \label{Simulations}

\subsection{Simulation Methods} \label{SimulationMethods} 

In this section, we evaluate the performance of the RBC method and the $D$ measure \eqref{Dmeasure} through simulation studies. We fit four different RBC models with the distributions on $\bm{u}$ given in Table \ref{Table:1}. We denote these models as RBC-normal,  RBC-Laplace, RBC-$t$, and RBC-slash, respectively. For RBC-$t$, we set the degrees of freedom $\nu$ as $\nu = 4$, and for RBC-slash, we set the shape parameter $\xi$ as $\xi = 1$. For all RBC models, we ran the MCMC algorithm for 20,000 iterations, discarding the first 10,000 samples as burn-in. We used the posterior samples to estimate the posterior mean $\widehat{\theta} = \mathbb{E} (\theta | \bm{y})$ and  construct 95\% credible intervals for $\theta$. Finally, for each of the RBC models, we computed the DIC as in \eqref{DIC}. We used the model with the smallest DIC to compute the $D$ measure.

We compared the RBC model to two frequentist Copas selection models that assume normality for the study-specific random effects:
\begin{itemize}
		\item SMA: the standard meta-analysis model \eqref{metaanalysismodel} that does not account for publication bias;
	\item Copas: the original (frequentist) Copas selection model \eqref{Copasmodel}; and
	\item CLS: the Copas-like selection model of \cite{NingChenPiao2017}.
\end{itemize}
The Copas selection model was implemented using the \textsf{R} function \texttt{copas} in the \textsf{R} package \texttt{metasens}, which uses a grid search to tune $(\gamma_0, \gamma_1)$. We bootstrapped the residuals to obtain an estimate of the standard error (s.e.) of $\theta$. The CLS model uses an EM algorithm to compute the MLEs for $(\theta, \tau, \rho, \gamma_0, \gamma_1)$ simultaneously.  To  estimate the s.e.'s for $\theta$ under CLS, we used the inverse Hessian matrix. For both Copas and CLS, we constructed the 95\% confidence intervals as $\widehat{\theta} \pm 1.96 \times \textrm{s.e.} (\widehat{\theta})$. The SMA method obtains MLEs for $(\theta, \tau)$ under model \eqref{metaanalysismodel} without accounting for publication bias. Similarly to CLS, we used the inverse Hessian matrix to estimate standard errors and 95\% confidence intervals for SMA. Functions to implement CLS and SMA are provided in the \textsf{R} package \texttt{RobustBayesianCopas}.

For our synthetic experiments, we simulated a meta-analysis of $n=30$ studies under the model \eqref{Copasmodel}. We generated the within-study standard errors $s_i, i=1, \ldots, n$, from $\mathcal{U}(0.2, 0.8)$, and we set $\theta = 0.4$, $\tau = 0.2$, $\gamma_0 = -1$, and $\gamma_1 = 0.3$. We varied the correlation $\rho \in \{ 0.3, 0.6, 0.9 \}$, so that we could evaluate the methods in Section \ref{SimulationMethods} under varying degrees of publication bias. Note that when publication bias is present, we usually expect $\rho$ to be positive, because studies with larger effect sizes are more likely to published. 

We considered four simulation settings for the distribution of the random effects $u_i, i=1, \ldots, n$, in \eqref{Copasmodel}:
\begin{itemize}
\item \textbf{Experiment 1} (heavy-tailed errors): $u_i \sim t_{3}$;
\item \textbf{Experiment 2} (standard normal errors): $u_i \sim \mathcal{N}(0,1)$;
\item \textbf{Experiment 3} (extreme outliers): $u_i \sim 0.85 \mathcal{N}(0, 0.2) + 0.15 \mathcal{N}(0,20)$; 
\item \textbf{Experiment 4} (skewed right): $u_i \sim 0.9 \mathcal{N}(0,1) + 0.1 \mathcal{N}(5.5, 0.5)$.
\end{itemize}
In Experiment 3, most of the random effects were very close to zero because of the first mixture component $\mathcal{N}(0,0.2)$. However, the second  mixture component $\mathcal{N}(0,20)$ ensured that there were a small number of extreme outliers in both directions. In Experiment 4, the mixture component $\mathcal{N}(5.5, 0.5)$ ensured that the $u_i$'s had long right tails due to a few large values to the right of zero.

For the four experimental settings, we implemented the seven models and computed the bias ($\widehat{\theta} - \theta$) and the coverage probability (CP), i.e. the proportion of times the 95\% posterior credible or confidence intervals contained $\theta$. For the RBC method, we computed the $D$ measure \eqref{Dmeasure} using the model that gave the lowest DIC. 

\begin{figure}[H]
	\centering
	\includegraphics[width=\linewidth]{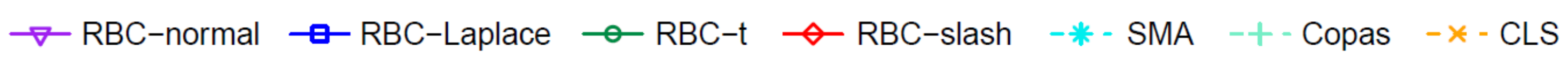}
	\includegraphics[width=.98\linewidth]{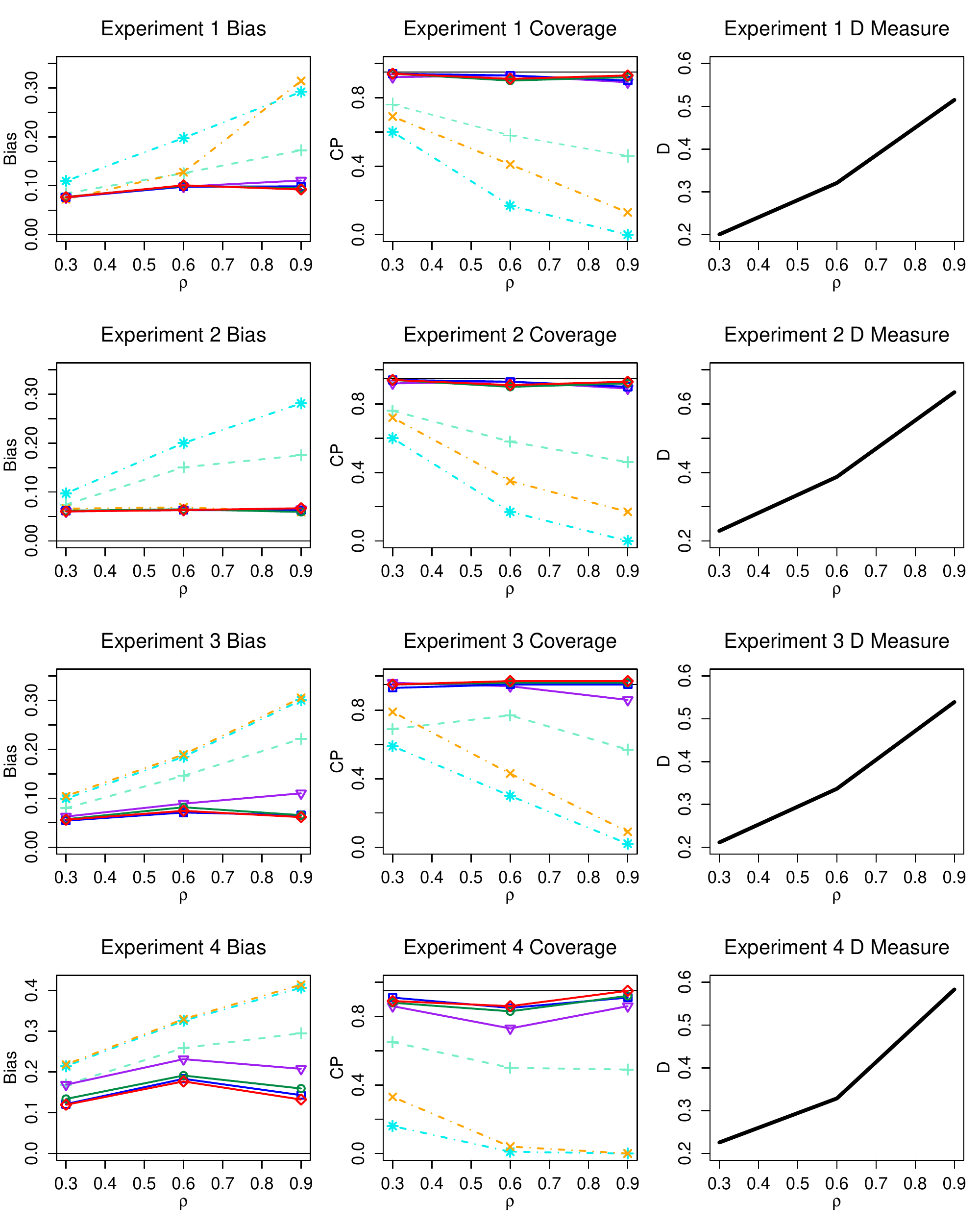}
	\caption{Our simulation results averaged across 100 replications. The left and center panels plot the bias and coverage for Experiment 1 (heavy-tailed random effects), Experiment 2 (standard normal random effects), Experiment 3 (extreme outliers), and Experiment 4 (skewed right) for the methods: RBC-normal ({\color{Orchid}$\bm{\bigtriangledown}$}), RBC-Laplace ({\color{blue}$\bm{\Box}$}), RBC-$t$ ({\color{OliveGreen}$\bm{\bigcirc}$}), RBC-slash ({\color{red}$\bm{\Diamond}$}), SMA ({\color{Cyan}$\bm{*}$}), Copas ({\color{Aquamarine}$\bm{+}$}), and CLS ({\color{orange}$\times$}). The right panel plots the $D$ measure for the RBC method which gave the lowest DIC.}
	\label{fig:experimentplots}
\end{figure}

\subsection{Simulation Results} \label{SimulationResults}

Figure \ref{fig:experimentplots} plots our simulation results for Experiments 1 through 4 described in Section \ref{SimulationMethods} averaged across 100 replications. The left and center panels of Figure \ref{fig:experimentplots} show the average bias and CP for Experiment 1 (heavy-tailed random effects), Experiment 2 (standard normal random effects), Experiment 3 (extreme outliers), and Experiment 4 (skewed right) for the seven methods we described in \ref{SimulationMethods}. To highlight the benefits of the Bayesian RBC approach vs. the frequentist ones in Figure \ref{fig:experimentplots}, we plotted the four RBC models with solid lines and the three frequentist approaches with dashed lines.

Figure \ref{fig:experimentplots} shows that in the experiments where there was departure from normality (Experiments 1, 3, and 4), RBC-Laplace, RBC-$t$, and RBC-slash all had lower bias and higher CP than the other four methods that assumed standard normal random effects. In the experiments where there were a few extreme outliers (Experiments 3 and 4), the RBC-Laplace, RBC-$t$, and RBC-slash models also achieved very close to the nominal coverage. This is because these heavy-tailed densities are much less sensitive to the presence of outliers and are more robust to violations of the normality assumption. It is interesting to note that even though the RBC-normal approach had higher bias and lower CP than the heavy-tailed RBC models in Experiments 1, 3, and 4, RBC-normal still had lower bias and much better coverage than SMA, Copas, or CLS. This suggests that for selection models, Bayesian posterior inference can often produce more accurate point estimates and more well-calibrated uncertainty quantification than frequentist approaches, even when the random effects distribution is misspecified.

When the study-specific effects truly were distributed as standard normal (Experiment 2), the CLS method performed equally as well as the RBC methods in terms of bias. However, the CP for CLS was lower. Moreover, in this scenario, \textit{all} four of the RBC models produced similar bias reduction and excellent coverage. This shows that the RBC-Laplace, the RBC-$t$, and RBC-slash models are also robust in the sense that they perform equally as well as RBC-normal when the normality assumption does hold.

Finally, the right panel of Figure \ref{fig:experimentplots} plots the $D$ measure \eqref{Dmeasure} for the RBC model which gave the lowest DIC against the correlation parameter $\rho$. We see that greater values of $\rho$ resulted in greater values of $D$. In particular, when $\rho$ was equal to 0.9, the average $D$ was above 0.5 in all four experiments, indicating the presence of high publication bias. Our simulation results illustrate that our proposed $D$ statistic is able to detect and quantify the magnitude of publication bias. 

\section{Real Data Applications} \label{DataApplication}

\subsection{Relationship Between Second-Hand Tobacco Smoke and Lung Cancer} \label{SmokingAnalysis}

We first applied the RBC model to a meta-analysis of studies on the relationship between second-hand tobacco smoke and lung cancer. \citet{HackshawLawWald1997} previously analyzed the results from 37 studies that evaluated the risk of developing lung cancer in women who were lifelong \textit{non}-smokers but whose husbands smoked, compared to women whose husbands had never smoked. In particular, \citet{HackshawLawWald1997} fit a random effects meta-analysis model \eqref{metaanalysismodel}, resulting in a pooled odds ratio (OR) of 1.24 and a 95\% confidence interval of (1.13, 1.36). \citet{HackshawLawWald1997} concluded that married, non-smoker women who were exposed to secondhand smoke by their smoker husbands were 24\% more likely to develop lung cancer than those whose husbands did not smoke. 

\begin{figure}[t!]
	\centering
	\includegraphics[width=0.495\linewidth]{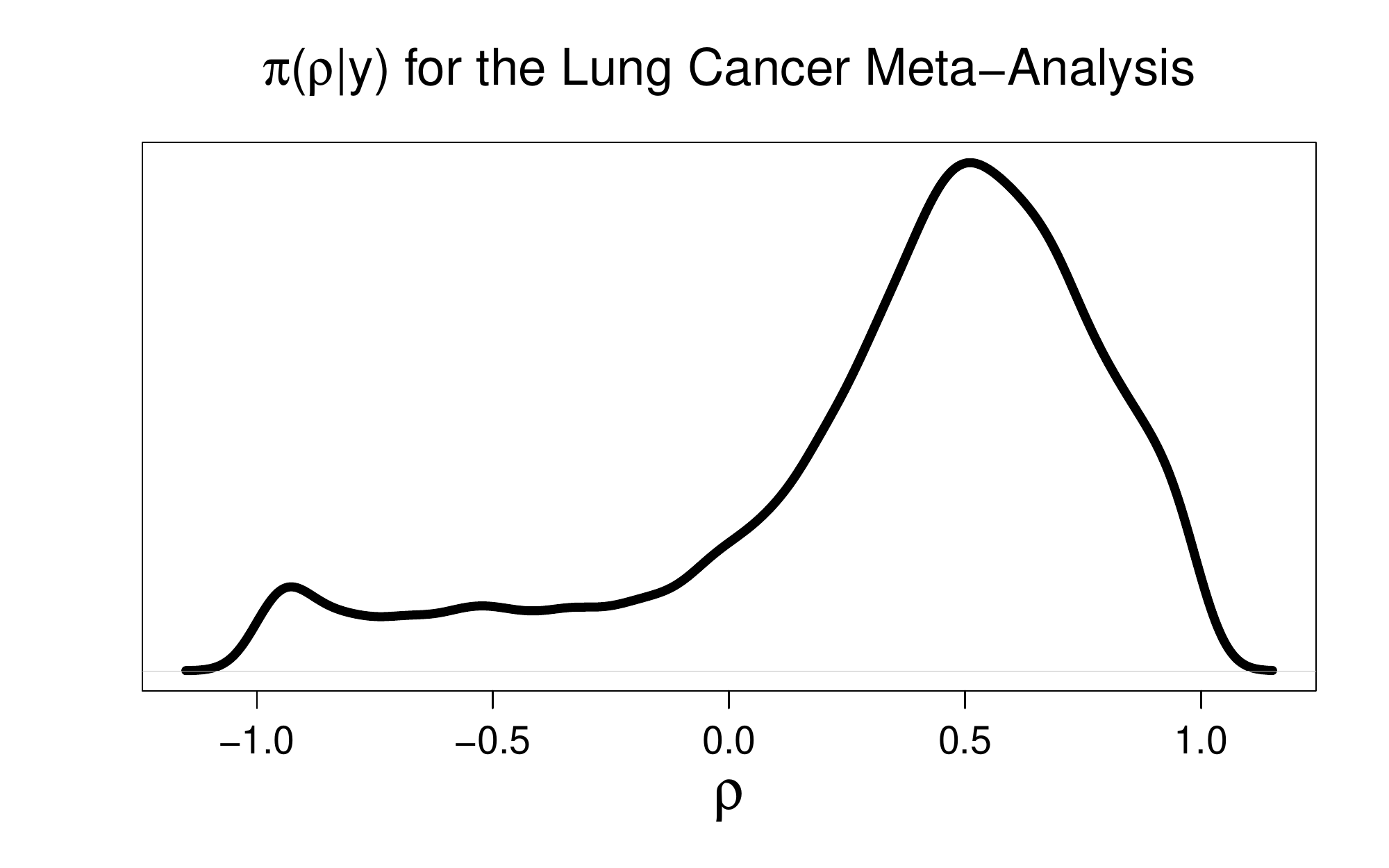}
	\includegraphics[width=0.495\linewidth]{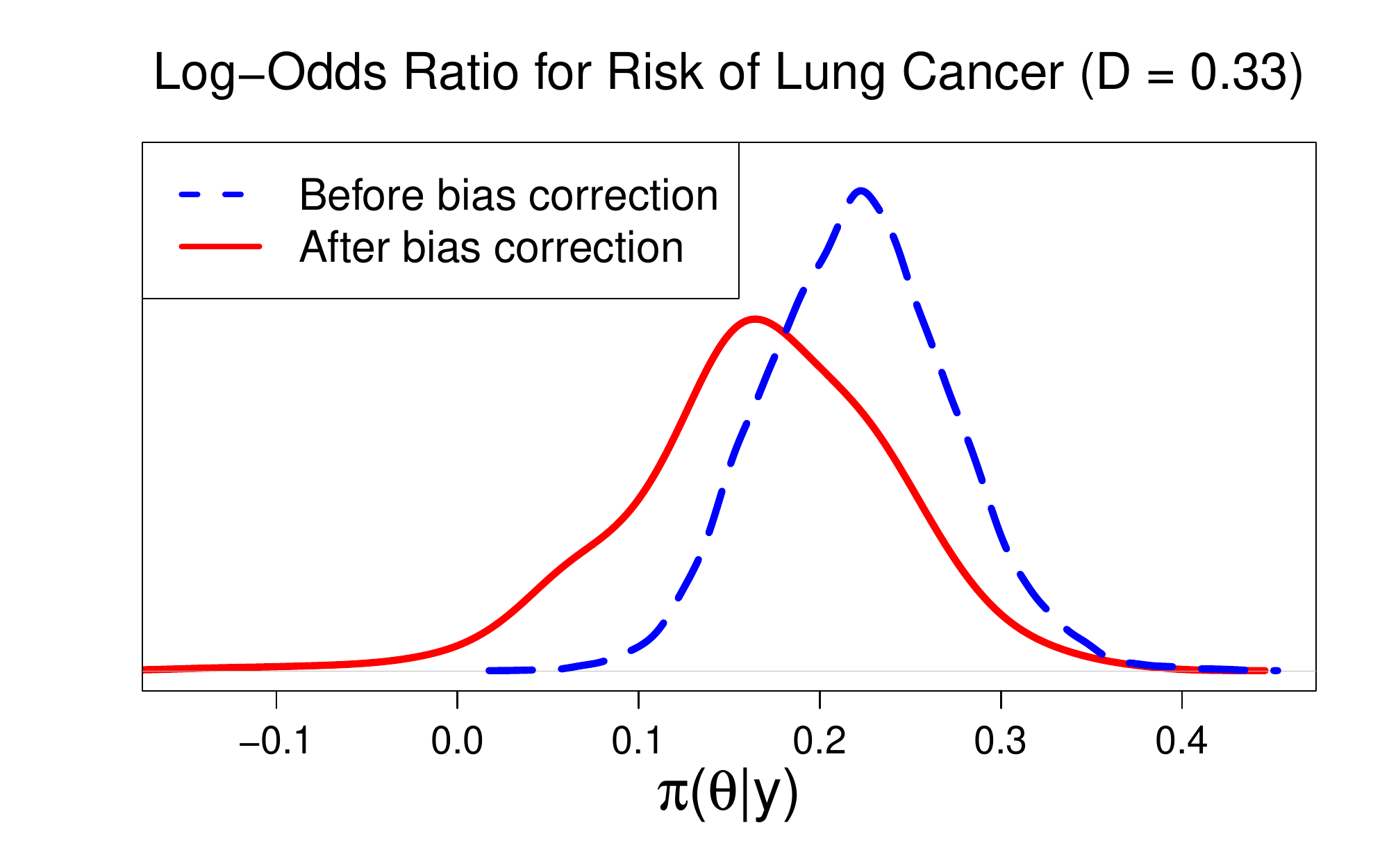}
		\includegraphics[width=0.495\linewidth]{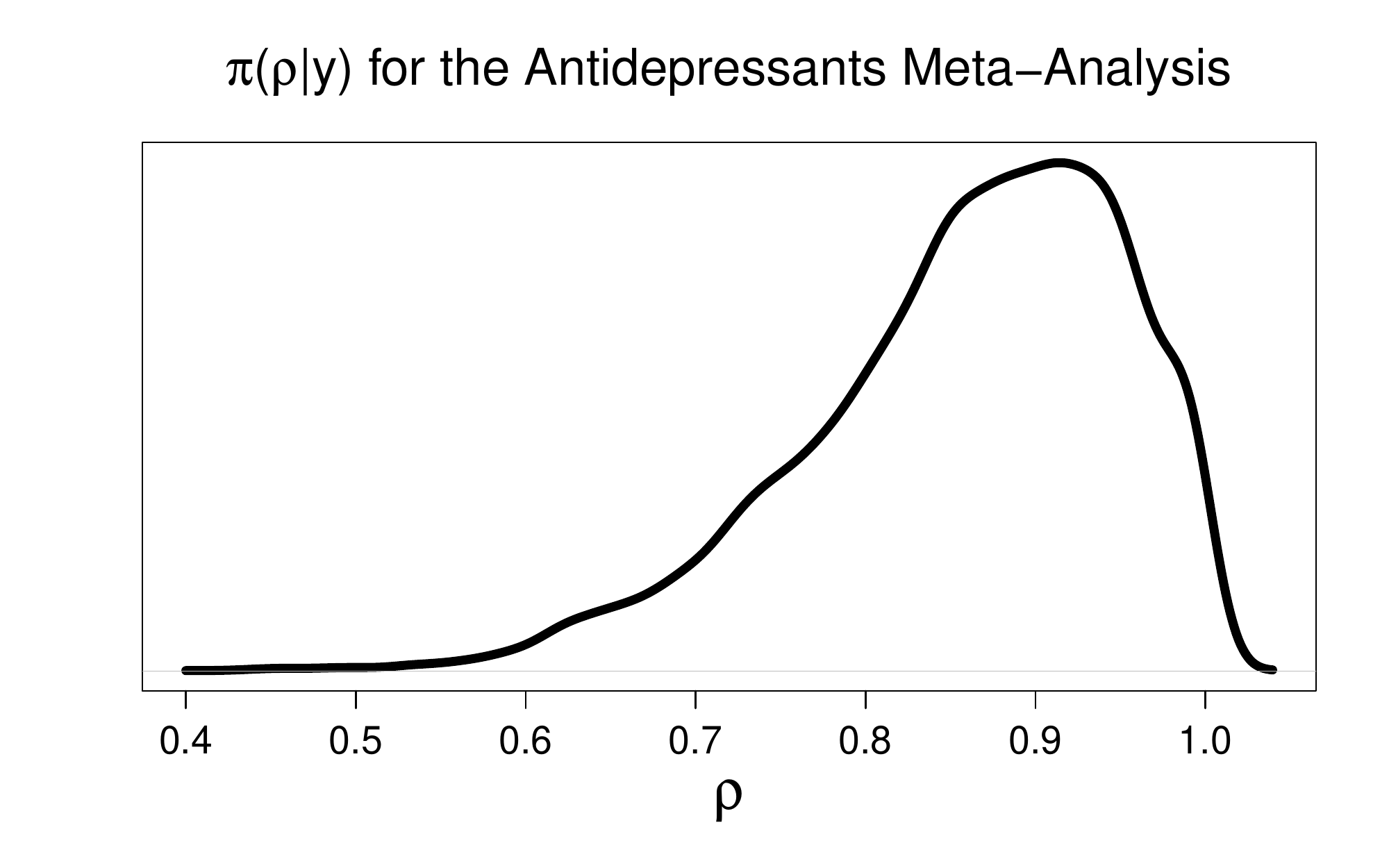}
	\includegraphics[width=0.495\linewidth]{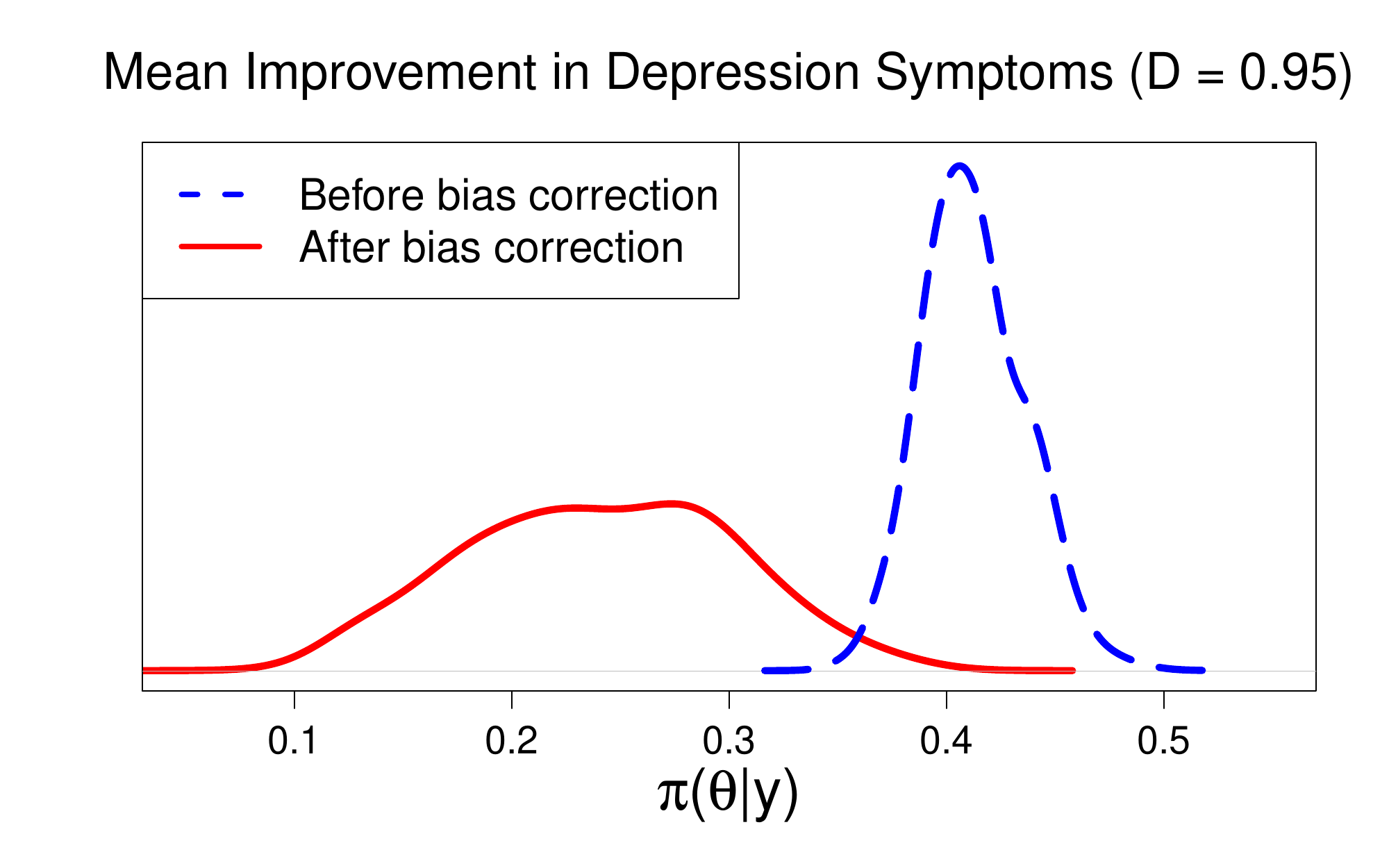}
	\caption{Results from our real data examples. The top-left panel plots the posterior distribution $\pi(\rho | \bm{y})$ for the meta-analysis on the risk of developing lung cancer from second-hand smoke. The top-right panel plots the posterior distributions for the log-odds ratio of developing lung cancer, $\pi_{rbc}( \theta | \bm{y})$ (solid line) and $\pi_{\rho=0} ( \theta | \bm{y})$ (dashed line). In the bottom-left panel, we plot the posterior distribution $\pi(\rho | \bm{y})$ for the meta-analysis on antidepressents. In the bottom-right panel, we plot the posterior distributions for the mean improvement in depression symptoms, $\pi_{rbc} (\theta | \bm{y})$ (solid line) vs. $\pi_{\rho = 0} (\theta | \bm{y})$ (dashed line).}
	\label{fig:smokingantidepressants}
 \end{figure}

 Previous analysis of this data by \citet{NingChenPiao2017} suggested some evidence of publication bias. Motivated by this possibility, we used our method to estimate $\theta$, the log-odds ratio for developing lung cancer. We fit four different RBC models with normal, Laplace, Student's $t$, and slash distributions for the random effects. The RBC model with $t$-distributed errors gave the lowest DIC, so we used this model to conduct our final analysis.
 
  We first performed inference about the presence of publication bias using the posterior distribution for $\pi(\rho | \bm{y})$. The top-left panel of Figure \ref{fig:smokingantidepressants} shows that $\pi(\rho | \bm{y})$ is concentrated about 0.5, with a posterior median of $\widehat{\rho} = 0.46$. This suggests the presence of publication bias. Note that for $\rho$, we use the posterior median as a point estimate rather than the mean, since the posterior for $\pi( \rho | \bm{y})$ is often skewed (and hence the posterior mean is heavily influenced by a few extreme values near $-1$ and $1$). While the posterior $\pi(\rho | \bm{y})$ is useful for assessing the presence of publication bias, it is not as informative as the $D$ measure in quantifying how much the posterior $\pi(\theta | \bm{y})$ changes \textit{once we have corrected} publication bias. 

 Next, we estimated $\theta$. The top-right panel of Figure \ref{fig:smokingantidepressants} displays the posterior distribution for $\pi_{rbc}( \theta| \bm{y})$ (solid line) against the posterior for $\pi_{\rho=0} ( \theta | \bm{y})$ (dashed line). We computed $D = 0.33$, which suggests a moderate magnitude of publication bias. In order to compare our results to those of \citet{HackshawLawWald1997}, we computed the posterior mean and 95\% posterior credible interval for the odds ratio $\exp(\theta)$. The RBC model gave a posterior mean OR of 1.19, with a 95\% credible interval of (1.02, 1.35).   In short, our analysis suggests that married, non-smoker women who were exposed to second-hand smoke by their husbands still had a significant risk of developing lung cancer, albeit a slightly lower risk than previously concluded (i.e. about 19\% more likely, as opposed to 24\% more likely \cite{HackshawLawWald1997}). 

\subsection{The Efficacy of Antidepressants} \label{DepressionAnalysis}

Although antidepressents are among the world's most widely prescribed drugs, there has been considerable controversy about their effectiveness. In 2008, \citet{TurnerMatthewsLinardatosRosenthal2008} presented a comparison of effectiveness data on depressants published in journals with the corresponding results from trials submitted to the Food and Drug Administration (FDA) between 1987 and 2004 for licensing. \citet{TurnerMatthewsLinardatosRosenthal2008} found evidence of bias towards results favoring active intervention. In particular, there were 73 studies with results as reported to the FDA (74 originally but two of them were subsequently combined), but only 50 (69\%) of these studies were subsequently published. 

We applied our method to a meta-analysis of these 50 published studies. In these studies, the outcome $\theta$ is a quantitative measure for improvement in depression symptoms. Since studies reported their outcomes on different scales, effect sizes were all expressed as standardized mean differences estimated by Hedges' $g$, accompanied by corresponding variances \cite{MorenoSuttonAdesCooperAbrams2011, Hedges1982}. This data set is available in the \textsf{R} package \texttt{RobustBayesianCopas}.

We fit four different RBC models with normal, Laplace, Student's $t$, and slash distributions for the random effects. The RBC model with normally distributed random effects gave the lowest DIC, so we used this model to conduct our final analysis. We first performed inference about $\rho$. The bottom-left panel of Figure \ref{fig:smokingantidepressants} displays the posterior distribution for $\pi(\rho | \bm{y})$. We see that $\pi(\rho | \bm{y})$ is highly concentrated on large values, with a posterior median of $\widehat{\rho} = 0.80$, which suggests very strong publication bias.

 Next, we considered estimates for $\theta$ under the RBC model, compared to those obtained from a standard meta-analysis \eqref{metaanalysismodel}. Under the standard model, we estimated the MLE $\widehat{\theta}_{mle} = 0.41$ with a 95\% confidence interval of $(0.36, 0.46)$. The posterior mean effectiveness under the RBC model, on the other hand, was only $\widehat{\theta}_{rbc} = 0.24$ with a 95\% posterior credible interval of $(0.13, 0.36)$.  Our results suggest that the mean improvement in depression symptoms from antidepressents may be lower than previously reported. 

 The bottom-right panel of Figure \ref{fig:smokingantidepressants} displays $\pi_{rbc} (\theta | \bm{y})$ (solid line) against $\pi_{\rho=0} (\theta | \bm{y})$ (dashed line). This plot shows that once we have corrected for the publication bias with the RBC model, we obtain significantly lower estimates of $\theta$ with greater uncertainty and very little overlap with the \textit{non}-bias-corrected posterior. We computed $D = 0.95$, indicating a very high magnitude of publication bias. The contrast between the two plots on the right in Figure \ref{fig:smokingantidepressants} shows that the $D$ measure is a very useful measure for quantifying uncertainty bias. 

\subsection{The Prevalence of Publication Bias}

The RBC model has shown promising performance in simulation studies and meta-analyses on real studies in Sections \ref{Simulations} and \ref{DataApplication}. Given its excellent empirical performance, we employ the RBC method to assess how prevalent the issue of publication bias is across multiple meta-analyses. Similarly, \citet{HigginsThompsonDeeksAltman2003} developed the $I^2$ statistic as a measure of consistency of the results of studies in meta-analyses. \citet{HigginsThompsonDeeksAltman2003} evaluated the performance of $I^2$ on 509 meta-analyses of dichotomous outcomes in the \textit{Cochrane Database of Systematic Reviews}. 

\begin{figure}[t!]
	\centering
	\includegraphics[width=0.85\linewidth]{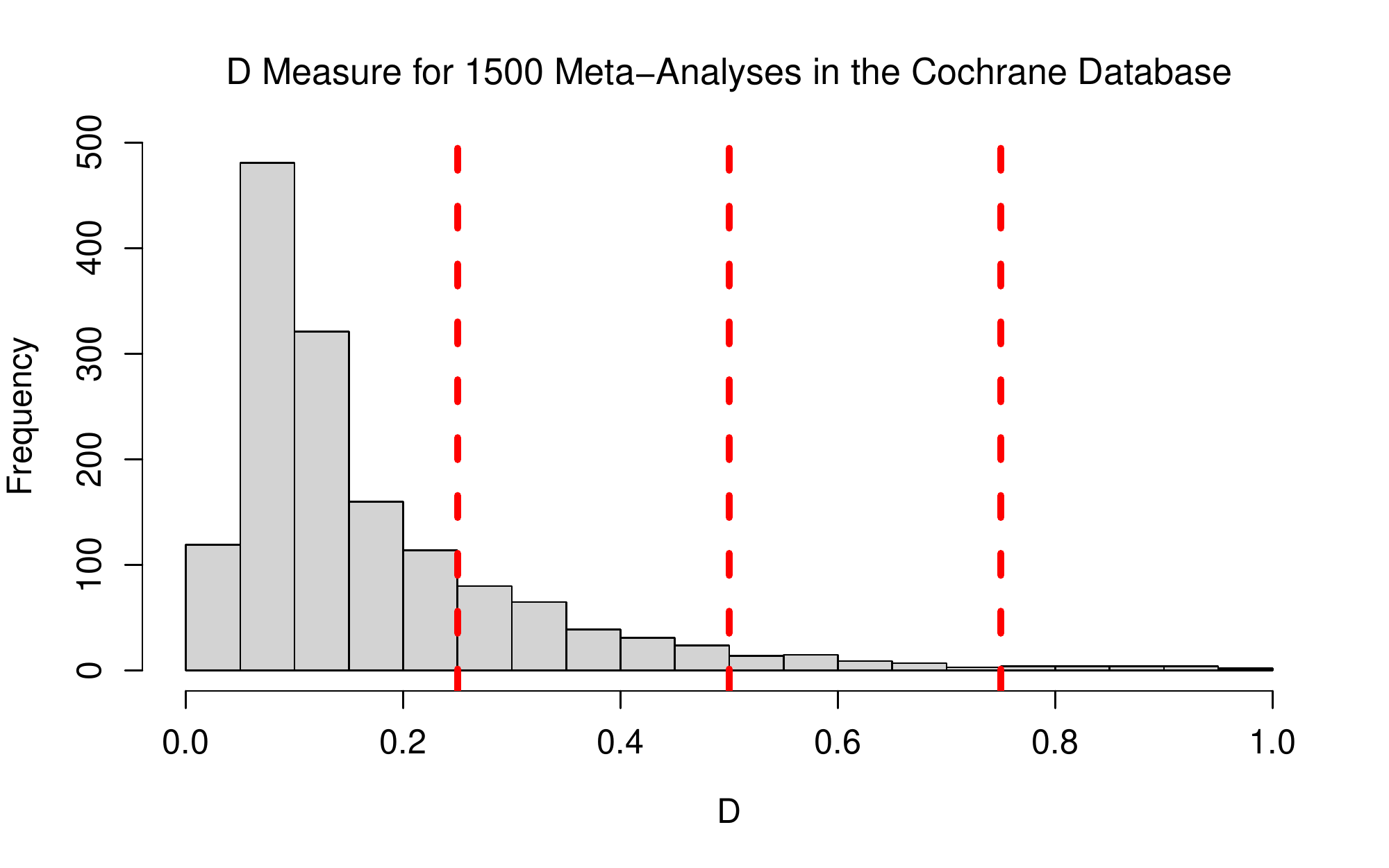}
	\caption{The empirical histogram of the $D$ measure evaluated for 1500 meta-analyses of dichotomous outcomes in the \textit{Cochrane Database of Systematic Reviews}. The dashed vertical lines correspond to the cutoffs ($D= 0.25$, 0.5, and 0.75) used to interpret the magnitude of publication bias.}
	\label{fig:DfsCochrane}
 \end{figure}

 Inspired by \citet{HigginsThompsonDeeksAltman2003}, we computed the $D$ measure for the overall treatment effect $\theta$ for a set of 1500 randomly selected meta-analyses of dichotomous outcomes from the \textit{Cochrane Database of Systematic Reviews} where each meta-analysis contained at least eight studies. As before, we fit four different RBC models with normal, Laplace, Student's $t$, and slash random effects and used the RBC model with the lowest DIC to compute the $D$ measure. The number of studies in these meta-analyses varied from $n=8$ to $n=135$. Figure \ref{fig:DfsCochrane} shows the empirical histogram for the $D$ measure. We found that 1195 (79.7\%) of these meta-analyses had negligible publication bias ($0 \leq D \leq 0.25$), 239 (15.9\%) had moderate publication bias ($0.25 < D \leq 0.5$), 48 (3.2\%) had high publication bias ($0.5 < D \leq 0.75)$, and 18 (1.2\%) had very high publication bias ($0.75 < D \leq 1$). 

\section{Discussion} \label{Discussion}

 In this paper, we have introduced the robust Bayesian Copas (RBC) selection model for correcting publication bias in meta-analysis. Our method can be implemented with heavy-tailed distributions for the random effects $\bm{u}$ in the Copas selection model \eqref{Copasmodel}, thereby providing robustness against deviations from normality for the study-specific effects. We also introduced the $D$ measure \eqref{Dmeasure} for quantifying the magnitude of publication bias. Specifically, $D$ quantifies the amount of dissimilarity between a standard random effects meta-analysis \eqref{metaanalysismodel} and a meta-analysis done with the Copas selection model \eqref{Copasmodel}. We illustrated that our method performs very well in a variety of simulation and real data settings.  We have provided an \textsf{R} package \texttt{RobustBayesianCopas} to implement our method, which can be found on the Comprehensive \textsf{R} Archive Network (CRAN). 

 In this paper, we have focused only on univariate meta-analysis. However, there is an increasing need to explore multivariate and network meta-analysis methods, which simultaneously analyze multiple treatments or comparisons between multiple treatments. See e.g. \cite{RileyJacksonSalantiBurkePriceKrkhamWhite2017, JacksonRileyWhite2011} for a review of different motivating applications and methods. For multivariate and network meta-analyses, the presence of publication bias will also lead to biased estimates of the treatment effects. In future work, we will extend the RBC model to multivariate and network settings.

\section*{Acknowledgments}
The bulk of this work was completed when the first author was a postdoc at the Perelman School of Medicine, University of Pennsylvania, under the supervision of the last two authors. Dr. Ray Bai and Dr. Mary Boland were funded in part by generous funding from the Perelman School of Medicine, University of Pennsylvania. Dr. Ray Bai and Dr. Yong Chen were funded by NIH grants 1R01AI130460 and 1R01LM012607. 

\bibliographystyle{chicago}
\bibliography{RobustBayesianCopasReferences}

\end{document}